\documentclass[epsf,usegraphicx]{mn2e}

\usepackage{xspace}
\usepackage{amsfonts}
\usepackage{pifont}

\title[Flares detected on the M dwarf star KIC 5474065]
{Short duration high amplitude flares detected on the M dwarf star KIC 5474065}

\author[]
{Gavin Ramsay$^{1}$, J. Gerry Doyle$^{1}$, Pasi Hakala$^{2}$, David Garcia-Alvarez$^{3,4,5}$,\and 
Adam Brooks$^{1,6}$  Thomas Barclay$^{7,8}$ Martin Still$^{7,8}$\\
$^{1}$Armagh Observatory, College Hill, Armagh, BT61 9DG\\
$^{2}$Finnish Centre for Astronomy with ESO (FINCA), University of Turku,
V\"{a}is\"{a}l\"{a}ntie 20, FI-21500 PIIKKI\"{O}, Finland\\
$^{3}$Instituto de Astrofísica de Canarias, E-38205 La Laguna, 
Tenerife Spain\\
$^{4}$Dpto. de Astrofísica, Universidad de La Laguna, 38206 La Laguna, Tenerife Spain\\
$^{5}$Grantecan CALP, 38712 Breña Baja, La Palma Spain\\
$^{6}$Mullard Space Science Laboratory, University College London,
Holmbury St. Mary, Dorking, Surrey, RH5 6NT\\
$^{7}$NASA Ames Research Center, M/S 244-40, Moffett Field, CA 94035, USA\\ 
$^{8}$Bay Area Environmental Research Institute, Inc., 560 Third St. West, 
Sonoma, CA 95476, USA\\}

\date{Accepted 2013 June 25.  Received 2013 June 25; in original form 2013 May 22}

\begin{document}
\outer\def\gtae {$\buildrel {\lower3pt\hbox{$>$}} \over 
{\lower2pt\hbox{$\sim$}} $}
\outer\def\ltae {$\buildrel {\lower3pt\hbox{$<$}} \over 
{\lower2pt\hbox{$\sim$}} $}
\newcommand{\ergscm} {ergs s$^{-1}$ cm$^{-2}$}
\newcommand{\ergss} {ergs s$^{-1}$}
\newcommand{\ergsd} {ergs s$^{-1}$ $d^{2}_{100}$}
\newcommand{\pcmsq} {cm$^{-2}$}
\newcommand{\ros} {\sl ROSAT}
\newcommand{\chan} {\sl Chandra}
\newcommand{\xmm} {\sl XMM-Newton}
\newcommand{\kep} {\sl Kepler}
\def\rchi{{${\chi}_{\nu}^{2}$}}
\newcommand{\Msun} {$M_{\odot}$}
\newcommand{\Mwd} {$M_{wd}$}
\newcommand{\Lsol} {$L_{\odot}$}
\def\Mdot{\hbox{$\dot M$}}
\def\mdot{\hbox{$\dot m$}}
\newcommand{\teff}{\ensuremath{T_{\mathrm{eff}}}\xspace}
\newcommand{\tickYes}{\checkmark}
\newcommand{\tickNo}{\hspace{1pt}\ding{55}}
\newcommand{\src} {KIC 5474065}
\newcommand{\srctwo} {KIC 9726699}

\maketitle

\begin{abstract}

Using data obtained during the {\sl RATS-Kepler} project we identified
one short duration flare in a 1 hour sequence of ground based
photometry of the dwarf star {\src}. Observations made using GTC show
it is a star with a M4 V spectral type. {\kep} observations made using
1 min sampling show that {\src} exhibits large amplitude
($\delta$F/F$>$0.4) optical flares which have a duration as short as
10 mins. We compare the energy distribution of flares from {\src} with
that of {\srctwo}, which has also been observed using 1 min sampling,
and ground based observations of other M dwarf stars in the
literature. We discuss the possible implications of these short
duration, relatively low energy flares would have on the atmosphere of
exo-planets orbiting in the habitable zone of these flare stars.

\end{abstract}


\section{Introduction}

Flares with duration of a few to tens of minutes and energies of
$\sim10^{28-35}$ ergs have been observed on low mass dwarf stars for
many decades (eg Bopp \& Moffett 1973, Gershberg \& Shakhovskaia
1983). The origin of these flares is thought to be similar to Solar
flares in that they are produced during magnetic reconnection events
(eg Haisch, Strong \& Rodono 1991). Studying stellar flares from a
wide range of stars can give important insight to how magnetic
activity varies as a function of stellar mass and age. In more recent
years, the affects of flares on the atmosphere of exo-planets around
dwarf stars has been the subject of much interest (eg Segura et al
2010).
 
Historically the study of stellar flares was performed on known M
dwarf stars. However, with the advent of large scale surveys such
  as SDSS it has become possible to identify events from many
  previously unknown flare stars (eg Davenport et al 2012).  Whilst
  this will no doubt prove a goldmine for stellar flare researchers,
  the issue of separating extra-galactic transient events and flares
  from M dwarfs will become increasingly difficult in future surveys
  such as that made using LSST.

One survey which allows the virtually uninterrupted observation of
sources is NASA's {\kep} mission which covers an area of 116 square
degrees. The light curves extend over many months (or years) and have
a precision of parts per million and allows models of stellar
structure to be tested in a way not previously possible. A key point
is that the actual targets which are observed using {\kep} can be
updated every month. Walkowicz et al (2011) presented {\kep}
observations of flares seen in cool stars, while Balona (2012)
reported observations of stars with A/F spectral type and Maehara
  et al (2012) presented some examples of `super' flares on Solar type
  stars.

In June 2011 we started the {\sl RATS-Kepler}\footnote{\sl RApid Temporal
  Survey-Kepler} project whose aim was to identify sources which
showed flux variations on short ($<$30 min) timescales (Ramsay et al
2013). We do this by taking a series of short exposures (20 sec) using
wide field cameras on telescopes such as the Isaac Newton Telescope on
La Palma on specific fields for one hour. Light curves of each object
are derived and variable sources identified. One variable source which
we identified was {\src} which showed a short duration ($<$20 min)
flare with an amplitude of 0.6 mag in the $g$ band. We were successful
in placing {\src} on the {\kep} 1 min sampling target list. This paper
presents the results of these {\kep} observations and a comparison of
the energy distribution of the flares with other low mass flare stars.

\section{KIC 5474065}

Although \src\ ($\alpha_{2000}$=19h 53m 02.3s,
$\delta_{2000}$=+40$^{\circ}$ 40$^{'}$ 34.6$^{''}$) is included in the
Kepler Input Catalog (Brown et al 2011), it does not have a measured
temperature or surface gravity. It is, however, included in the
Kepler-INT Survey (U=20.61, g=19.00, r=17.33, i=15.60; Greiss et al
2012a,b); the UBV survey of the Kepler field (B=18.79, V=18.07;
Everett, Howell \& Kinemuchi 2012) and also the 2MASS survey
(J=14.015, H=13.397, K=13.215; Skrutskie et al 2006). The optical
colours indicate a relatively late-type star. Since \src\ is variable
(due to its rotational modulation and flare activity) some degree of
caution is required when determining its colours unless it is known
that multi-band observations are made simultaneously. However,
L\'{e}pine \& Gaidos (2011) show the relationship between the colour
$(V-J)$ and spectral type for late-type stars.  For \src, $V-J$ = 4.06
implies a spectral type of M3 to M4.

\section{Gran Telescopio Canarias Spectroscopic Data}

We carried out low-resolution spectroscopy with the Optical System for
Imaging and Low Resolution Integrated Spectroscopy (OSIRIS) tunable
imager and spectrograph (Cepa et~al. 2003) at the 10.4\,m Gran
Telescopio Canarias (GTC), located at the Observatorio Roque de los
Muchachos in La Palma, Canary Islands, Spain.  The heart of OSIRIS is
a mosaic of two 4k\,$\times$\,2k e2v CCD44--82 detectors that gives an
unvignetted field of view of 7.8\,$\times$\,7.8\,arcmin$^{2}$ with a
plate scale of 0.127\,arcsec\,pix$^{-1}$.  However, to increase the
signal-to-noise ratio of our observations, we chose the standard
operation mode of the instrument, which is a 2\,$\times$\,2-binning
mode with a readout speed of 100\,kHz.

Two spectra each with an exposure of 300 sec were obtained using the
OSIRIS R1000R grism in service mode on 11 May 2013. They were made as
part of a GTC filler programme which utilies poor weather
conditions. We used the 1.0$^{''}$-width slit, oriented at the
parallactic angle to minimise losses due to atmospheric dispersion.
The resulting resolution, measured on arc lines, was R $\sim$ 700 in
the approximate 5250--9200\,{\AA} spectral range. The star Ross 640
was used to remove the instrumental response. The data were reduced
using standard {\tt Figaro}
routines\footnote{http://starlink.jach.hawaii.edu}.

\begin{figure}
\begin{center}
\setlength{\unitlength}{1cm}
\begin{picture}(8,5.5)
\put(0,0){\includegraphics{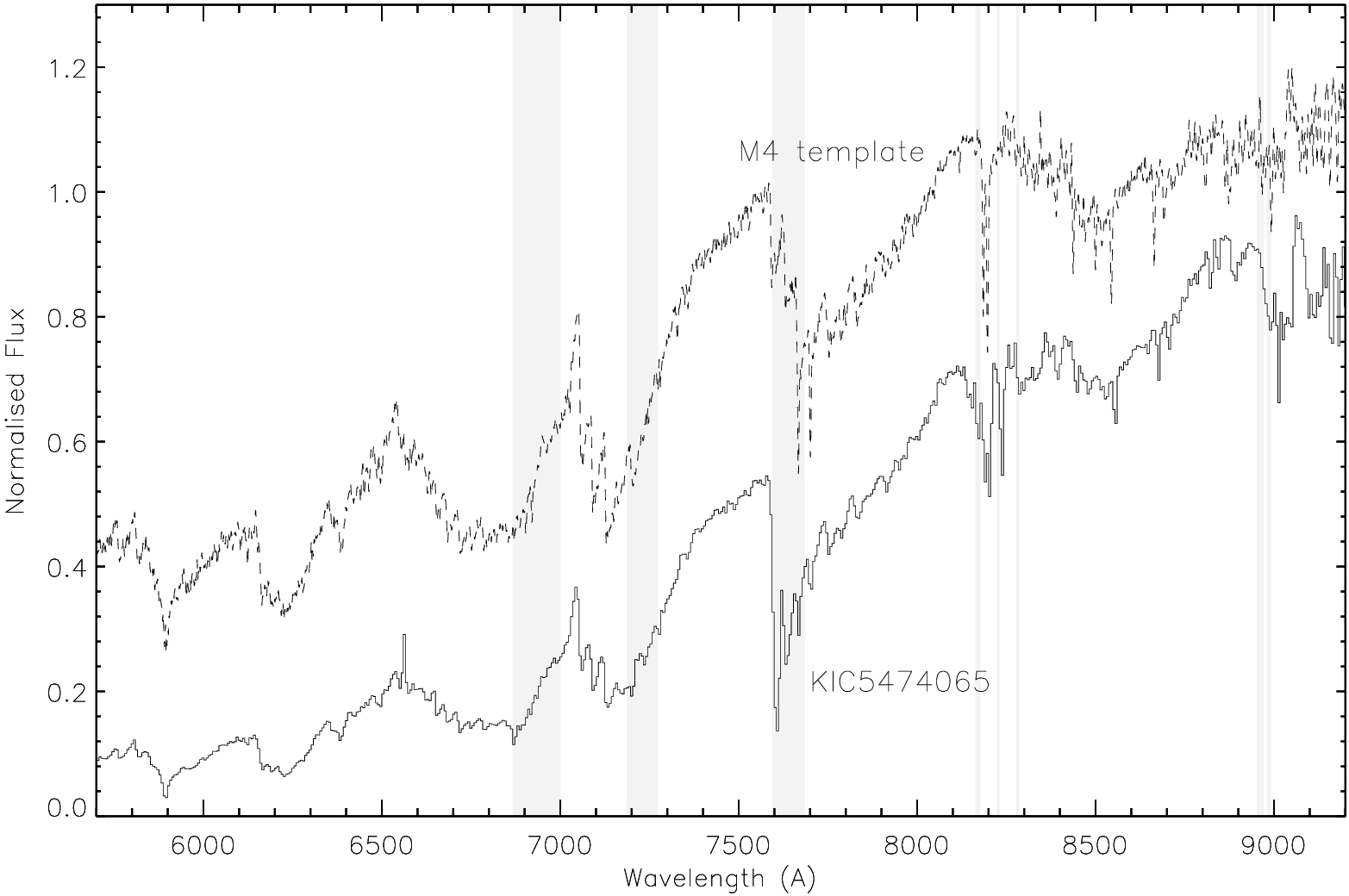}}
\end{picture}
\end{center}
\caption{The lower spectrum (solid line) shows the optical spectrum of
  {\src} obtained using GTC and Osiris. We indicate the
    wavelength of telluric absorption features as light grey vertical
    bands (taken from Kirkpatrick, Henry \& McCarthy 1991) and we have
    not attempted to remove them. The upper spectrum (dashed line)
    shows the M4 template spectrum provided by Bochanski et
    al. (2007).}
\label{spectrum}
\end{figure}

We show the optical spectrum of \src\ in Figure \ref{spectrum}: it is
clearly a late-type dwarf star. Examining Figure 1 of Bochanski et al
(2007), \src\ is later than an M0V spectral type. Judging by the depth
of the Na~{\sc i}~(8190\AA) feature and the Ca~{\sc ii} triplet around
8500 \AA\ it is most likely that \src\ has a M4V spectral type
although M3V and M5V are also possible.

In order to determine the energies of the flares, we must first
estimate the intrinsic luminosity of \src. L\'{e}pine \& Gaidos (2011)
include $VJHK$ and parallax data for late type stars. We were able to
extract data as a function of spectral type and estimate the mean
$M_{V}$ using relationship between $(V-J)$ and $M_{V}$ outlined in
L\'{e}pine \& Gaidos (2011). We show in Table \ref{Mv} the mean
absolute $V$ magnitude for spectral types M3V--M5V and we assume the
Sun has $M_{V}$ = 4.83 and $L_{\odot}=3.8\times10^{33}$ erg s$^{-1}$.

\begin{table}
\caption{The absolute magnitude and luminosity of stars with spectral
  types M3 V -- M5 V based on the data in L\'{e}pine \& Gaidos (2011).}
\begin{center}
\begin{tabular}{lrr}
\hline
Spectral Type & $M_{V}$ & L (erg s$^{-1}$) \\
\hline
M3 V & 11.2 & $1.1\times10^{31}$\\
M4 V & 12.4 & $3.6\times10^{30}$\\
M5 V & 13.5 & $1.4\times10^{30}$\\
\hline
\end{tabular}
\end{center}

\label{Mv}
\end{table}

\section{Kepler Data}

The detector on board {\kep} is a shutterless photometer using 6 sec
integrations and a 0.5 sec readout. There are two modes of
observation: {\it long cadence} (LC), where 270 integrations are
summed for an effective 28.4 min exposure, and {\it short cadence}
(SC), where 9 integrations are summed for an effective 58.8 sec
exposure.  When an object is observed in SC mode, LC data is also
automatically recorded. 

\begin{table}
\caption{The start and end times of the Short Cadence and Long 
Cadence {\kep} observations of \src.}
\begin{center}
\begin{tabular}{lrr}
\hline
Mode & Start & End \\
\hline
SC & 2012-06-28 15:03:34  & 2012-07-29 12:02:49  \\
LC & 2012-06-28 15:17:47  & 2012-10-03 19:40:10  \\
\hline
\end{tabular}
\end{center}
\label{log}
\end{table}

\begin{figure*}
\begin{center}
\setlength{\unitlength}{1cm}
\begin{picture}(16,11)
\put(-0.5,-0.5){\includegraphics{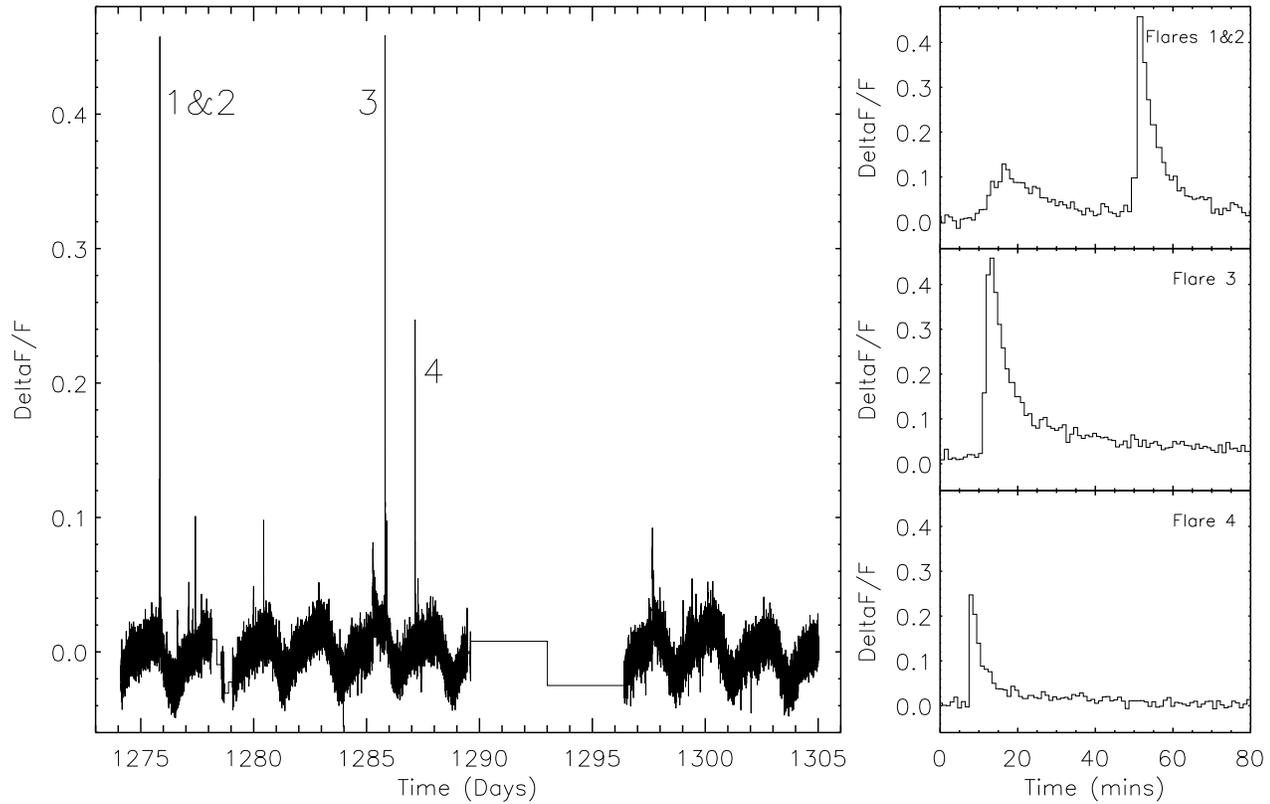}}
\end{picture}
\end{center}
\caption{The {\kep} Short Cadence light curve of {\src} made in Q14.
  $\Delta$F/F is the ratio of the difference between the flux at any
  point and the mean flux. The rotational period of 2.47 days is
  clearly seen as are the short duration flares. In the left hand panel
we zoom in on four flares.}
\label{light}
\end{figure*}

{\src} was observed using {\kep} in SC mode in Quarter 14: the start
and end times of the observations are shown in Table \ref{log}. The
on-time for SC mode was $\sim$24.1 days and 75 days for LC mode. After
the data are corrected for bias, shutterless readout smear and sky
background, light curves are extracted using simple aperture
photometry (SAP). Data which do not conform to `SAP\_QUALITY=0' were
removed (for instance, time intervals of enhanced solar activity) and
the data were corrected for systematic trends.

The light curve of \src, shown in Figure \ref{light}, exhibits two
main features -- one a clear quasi-sinusodial modulation with a period
of 2.47 days and a semi-amplitude of $\sim$2 percent, and the presence
of short but intense flares: two flares have an intensity $\Delta
F/F\sim$46 percent. For comparison, {\sl MOST} observations of the
dM3e star AD Leo (made using a 1 min cadence), show flares with
$\Delta F/F\sim$28 percent (Hunt-Walker et al 2012), while
`super-flares' with amplitude of 8 percent are seen on Solar-type
stars (Maehara et al 2012). The rotational period of {\src} is typical
of M4-5 dwarfs, e.g. YZ CMI has a rotational period of 2.78 days,
while V577 Mon has a rotational period of 1.95 days.

For comparison, we also extracted the light curve of {\srctwo} 
  ($\alpha_{2000}$=19h 51m 09.4s, $\delta_{2000}$=+46$^{\circ}$
  29$^{'}$ 01.2$^{''}$) which has also been observed using {\kep} in
SC mode. Savanov \& Dmitrienko (2011) presented an analysis of this
data, but concentrated on determining the extent and duration and
spots on its photosphere and did not discuss the flares
themselves. Like {\src} it has a M4 V spectral type (Reid et al 2004),
but it is more rapidly rotating (a rotation period of 0.593 days)
which makes it similar to V374 Peg (0.45 days). {The {\sl Kepler
    Input Catlog} (Brown et al 2011) gives $g$=13.9 for {\srctwo}
  making it more than 5 mag brighter than {\src} and hence the {\kep}
  data of this source has a much higher signal to noise than {\src}.}

\begin{figure*}
\begin{center}
\setlength{\unitlength}{1cm}
\begin{picture}(16,8)
\put(0.5,-3.){\includegraphics{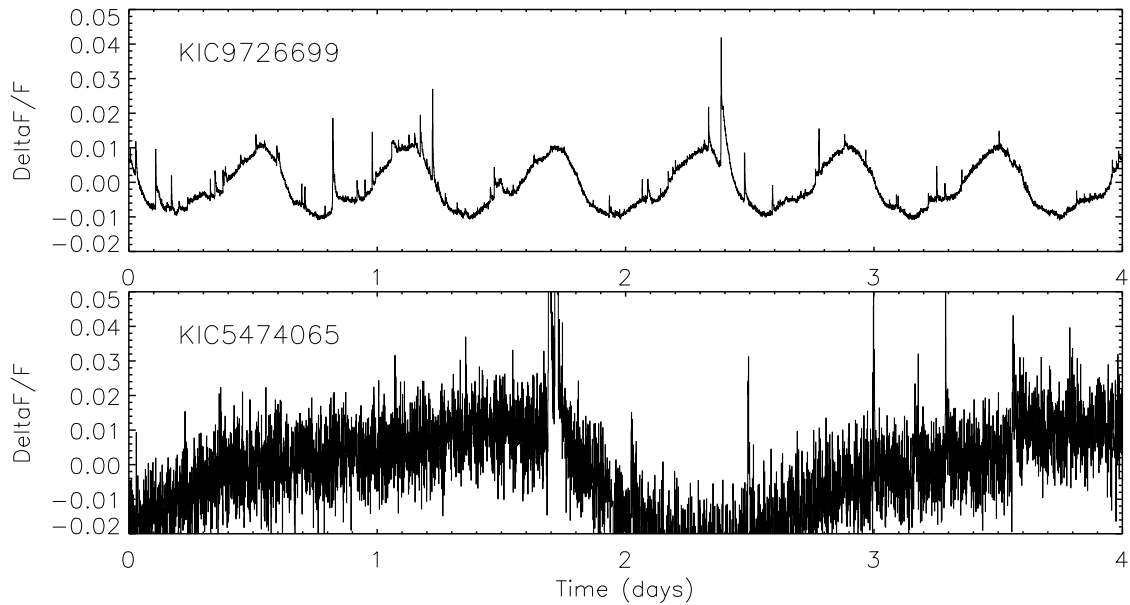}}
\end{picture}
\end{center}
\caption{In the top panel we show a section of the {\kep} Short
  Cadence light curve of KIC 9726699 made in Q6. $\Delta$F/F is the
  ratio of the difference between the flux at any point and the mean
  flux. For comparison we show in the lower panel a section of
    the light curve of {\src} on the same scales which is 5 mag
    fainter than {\srctwo}.}
\label{kic9726699}
\end{figure*}

{\srctwo} has been observed using {\kep} in SC mode in four quarters,
but here we have restricted our analysis to data from Q6.  We show a
4 day section of the light curve of {\srctwo} in Figure
  \ref{kic9726699}. It shows a relatively small number of large
amplitude flares, but its light curve is dominated by short duration,
low intensity, flares. For comparison, we show the light curve of
  {\src} also covering 4 days and on the same flux scale in Figure
  \ref{kic9726699}. Given that {\srctwo} is very much brighter
  compared to {\src}, it is possible that short duration low intensity
  flares are likely to be hidden in the noise in {\src}.

\section{Flare Characteristics}

To identify flares from {\src} in an automatic manner we first removed
the effects of the rotational modulation. After some experiment, we
identified the time interval when a flare occured when
($f_{i}-\overline{f})/\sigma>$3 where $f_{i}$ was the flux of the
$ith$ point, $\overline{f}$ was the overall mean of the light curve
and $\sigma$ was the standard deviation of the overall light
curve. The resulting flare times were then manually inspected and
  points which were clearly part of the same flare were edited to
  ensure no `double counting' of flares were made.  (If we defined a
lower threshold for flare detection the flare rate goes up but the
false positive rate also goes up as it becomes difficult to
distinguish between a genuine flare and noise in the data).  This
strategy found 27 flares in the SC light curve of {\src} -- in
otherwords, on average one flare was detected every 0.9 days, and on
average there was one flare with an intensity $\Delta$F/F$>$0.2 every
8 days.  There was no evidence of any pre-flare dips such as that seen
in V1054 Oph (Ventura et al 1995).  For {\srctwo} it was more
difficult to fully remove the effects of rotation and we therefore set
the detection threshold as ($f_{i}-\overline{f})/\sigma>$8. However,
we identified over 260 flares in the Q6 data of {\srctwo}. The
  fact that {\srctwo} appears to be more active compared to {\src} is
  consistent with the well known correlation between rotation period
  and stellar activity, (eg Noyes 1985), although we note that short
  duration, low energetic flares would not have been detected in
  {\src} due to the much higher noise level.

\begin{figure}
\begin{center}
\setlength{\unitlength}{1cm}
\begin{picture}(16,5.5)
\put(-0.5,0){\includegraphics{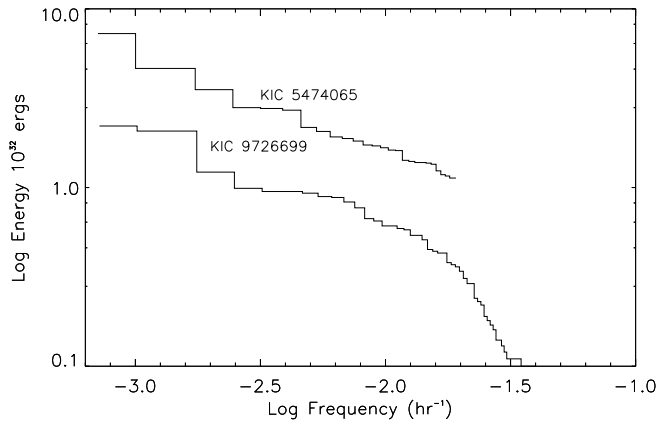}}
\end{picture}
\end{center}
\caption{The cumulative energy distribution of flares (in the {\kep}
  band-pass) as seen in {\src} and KIC 9726699.}
\label{cumulative}
\end{figure}

To derive the flare frequency rate, we use the following formulae
where $E=\sum_{f}$ is known as the flare equivalent duration, see Lacy
et al. (1976) for further details;

\begin{equation}
E=\sum_{f} [(I_{f+o}-I_{o})/I_{o}] \Delta T
\end{equation}

\noindent
where $I_{o}$ is the stellar intensity of the star in its quiescent
state, $I_{f+o}$ the intensity during a flare and $\Delta$T the
integration time. Further, we assumed the underlying luminosity was
$3.6\times10^{30}$ erg/s (Table \ref{Mv}) which is appropriate for a
M4V spectral type. This gave for {\src} a range in flare energy,
$L=1.1-7.3\times10^{32}$ ergs, whilst for {\srctwo} the range was
$L=0.01-2.2\times10^{32}$ ergs (these energies will vary by a
factor of three for one spectral sub-class either side of M4V).

\begin{figure*}
\begin{center}
\setlength{\unitlength}{1cm}
\begin{picture}(8,10)
\put(-4,0.5){\includegraphics{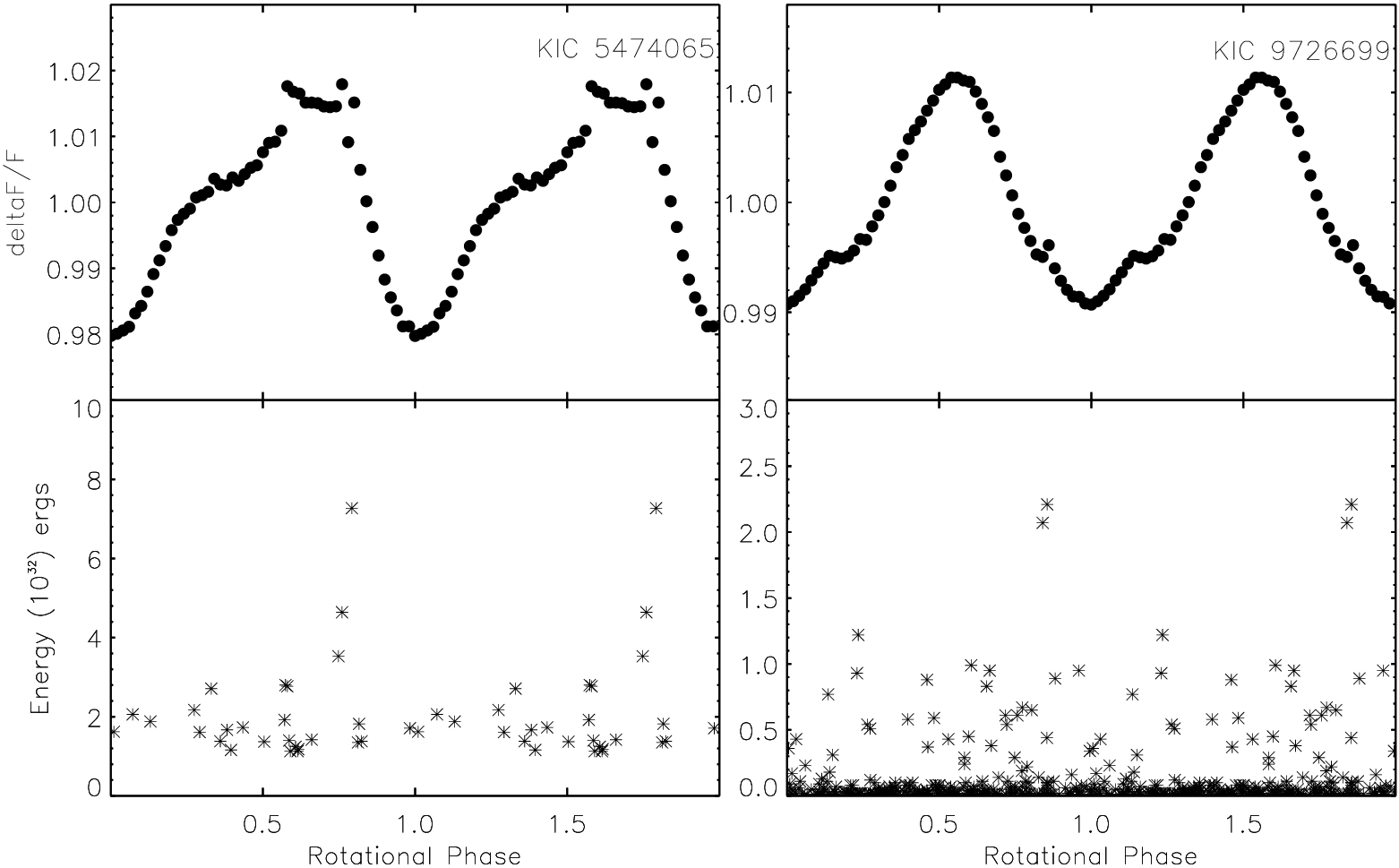}}
\end{picture}
\end{center}
\caption{In the upper panels we show the mean light curve of {\src}
  (left) folded on $T_{o}$= BJD 2456106.97 + 2.47295492 E and
  {\srctwo} (right) folded on $T_{o}$= BJD 2455371.310 + 0.593 E,
  where the light curves have been normalised so that the mean flux is
  unity and $\phi$=0.0 is defined as photometric minimum and the
    light curves have been binned into 50 bins. In the lower panels
  we show the energy emitted in each flare as a function of rotational
  phase. Deviations from a smooth curve are a result of flares in
    the light curve.}
\label{lum-phase}
\end{figure*}

Since the activity of flare stars is frequently measured in the $U$
band, we have estimated the equivalent energy of \src\ in the $U$
band. van den Oord et al (1996) found $E_{opt}/E_{U}=2.4$ where
$E_{opt}$ and $E_{U}$ is the energy emitted in the optical $UBV$ and
$U$ band respectively. In comparison, Lacy, Moffett \& Evans (1976)
find $E_{opt}/E_{WL}=2.1$ where $E_{WL}$ is the energy emitted in
'white light'. For our purposes we assume $E_{Kepler}/E_{U}=2.4$ and
note there will be a $\sim$10 percent uncertainty in our values of
$E_{U}$. For \src\ we therefore find the flares have a range between
$L_{U}=0.5-3\times10^{32}$ ergs and $L_{U}=0.01-0.9\times10^{32}$ ergs
for {\srctwo}.  The energy and duration of these flares are at the
lower energy of those detected on another dM4.5 flare star, AD Leo
(Pettersen et al. 1986).

We now show the cumulative flare-frequency distribution of the flares
in Figure \ref{cumulative}. This shows that on average {\src}
({\srctwo}) shows a flare with energy $L=10^{31}$ ergs every 0.2 days
(0.6 days) and a flare with energy $L=10^{32}$ ergs every 8.7 days
($\sim$117 days), compared to AD Leo which can produce a $10^{32}$
ergs flare every 1.5 days (see Table \ref{flare-rate}).

We show in Figure \ref{lum-phase} the light curve of both {\src} and
{\srctwo} folded on the stars rotational period together with the
energy of each flare as a function of the rotational phase. Since the
modulation in the light curve is caused by the rotation of stellar
spots into and out from view, the minimum flux corresponds to the
viewing phase when the fractional surface area covered by spots is at
its greatest (since the spots are cooler than the surrounding
photosphere).

In the case of {\src}, the three most energetic flares were seen in a
very short phase interval, $\phi$=0.74--0.79. Curiously, in {\srctwo}
the two most energetic flares were also seen at a similar phase
($\phi$=0.84--0.85) and separated by 6 rotational cycles. This
indicates that the most active regions on the star are preferentially
located in stellar longitude and last for timescales of (at least)
several weeks (a conclusion also reached by Savanov \& Dmitrienko
2010). In both sources, flares were seen at all rotational phases.

\section{{\src} and {\srctwo} as flare stars}

\begin{figure*}
\begin{center}
\setlength{\unitlength}{1cm}
\begin{picture}(16,10)
\put(0,11){\includegraphics{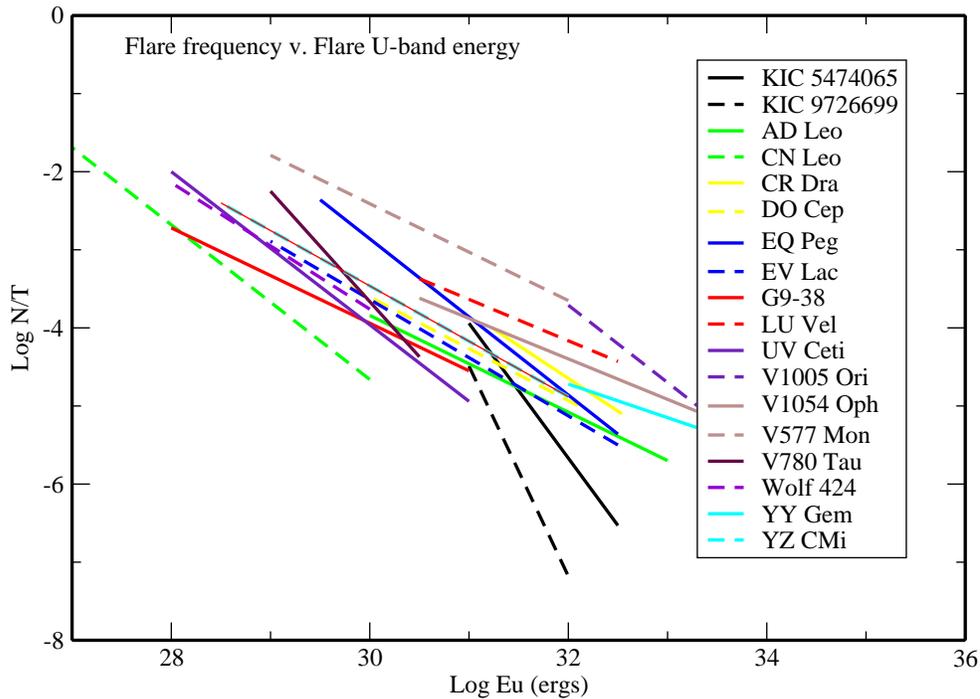}}
\end{picture}
\end{center}
\caption{The cumulative flare frequency (in seconds) versus U-band
  flare energy (in ergs) for a large group of M dwarfs plus that for
  KIC 5464065 and KIC 9726699. This data has been complied from work by Moffett
  (1974), Lacy et al.  (1976), Byrne et al. (1984, 1985), Pettersen
  (1975a,b, 1981a,b, 1983, 1985a,b, 2006a,b), Pettersen et al. (1983,
  1984, 1986), Pettersen \& Sundland (1991), Doyle \& Byrne (1986),
  Doyle et al. (1986, 1989, 1990), Hawley et al. (1989), Leto et
  al. (1997), Dal \& Evren (2010, 2011) and Dal (2011, 2012).}
\label{mdwarfs}
\end{figure*}

The monitoring of flares from M dwarf stars has been ongoing for the
past 50 years, most of it in the Johnson $U$-band.  Although all
  M dwarfs monitored over an extended timescale appear to show flares,
  only a few dozen have a well established flare-rate. In Figure
  \ref{mdwarfs} we summarise the results from several thousand hours
  of photometric monitoring (the caption indicates the original source
  of the data). Compared to the other M dwarf stars (which range from
dM0 to dM8) shown in Figure \ref{mdwarfs}, KIC 5474065 and KIC 9726699
show flares which are relatively energetic but occur less frequently.

{\srctwo} also produces more frequent but less energetic flares
  than {\src}, despite the fact that both have a dM4 spectral
  class. Generally speaking, stars of spectral class M4 and later are
  fully convective and therefore have a very different magnetic
  topology compared to stars with earlier spectral type. However,
  Morin et al. (2010) has shown this is not always the case as age may
  play a role in addition to mass and rotation period. V374 Peg has a
  similar rotation period as KIC 9726699, yet it can produce extremely
  energetic flares, e.g. Batyrshinova \& Ibragimov (2001) detected an
  11 mag superflare with an energy in excess of 10$^{35}$ erg. A more
  plausible explanation may be the relative spot coverage on these two
  stars. For instance Notsu et al. (2013) found that the energy of
  superflares is related to the total coverage of starspots and
  therefore the amount of magnetic energy stored around starspots.

For those stars where monitoring exists over a number of years, the
observed seasonably variability can be a factor of two, perhaps
indicating cycles similar to the Sun.  This has been determined in a
number of ways but includes narrow band photometric filters centered
on the Ca~{\sc ii} H \& K lines (eg Baliunas et al 1995) and
spectropolarimetric observations (eg Donati et al 2008) spread over a
considerable time interval. With the possibility of observing flare
stars using {\kep} with a cadence of 1 min for weeks at a time, it
will be practical to map the activity of many stars over a timescale
of years. This will also provide good motivation to re-examine the
effects that stars which show many flares have on the chemistry of
atmosphere's of exo-planets in the stars habitable zone.

\section{The impact of flares in the immediate stellar environment}

The implications of stellar flares on the atmosphere of an
  exo-planet orbiting around a flare star are important for the
  development of life as energetic flares could have a potentially
  hazardous influence on its habitability. For a M4 V dwarf star the
  habitable zone is roughly 0.04--0.14 AU (eg Kopparapu et al 2013). For
  instance Segura et al (2010) determined the potential effect of a
  flare as seen on the dM3e star AD Leo in April 1985. This flare
  which had a duration of 4 hrs was found to have an energy of
  $L\sim10^{34}$ ergs in the UV/Optical wave-band (Hawley \& Petterson
  1991), more than one order of magnitude than the largest flare seen
  on KIC 5474065. Segura et al (2010) determined that such a flare was
  not a direct hazard for life (as we know it) on an exo-planet 0.16
  AU distant from AD Leo. The flares which we report here are both
less energetic but also of much shorter duration. For {\src} the total
radiated $U$-band flare energy budget during the SC monitoring
interval was $\approx 10^{27}$ erg s$^{-1}$. Assuming that the total
radiated energy over all wavelengths is one order of magnitude greater
(eg Doyle \& Butler 1985) implies a value two orders of magnitude less
radiated energy than the large AD Leo flare referenced above which may
suggest a minimum effect on any nearby planet.  However, what effect a
sequence of frequently occurring flares still needs to be
investigated.

\begin{table}
\caption{Flare rates for KIC 5474065, KIC 9726699 compared to AD
  Leo where the energies are the equivalent energy in the $U$ band.}
\begin{center}
\begin{tabular}{llll}
\hline
                    & AD Leo            & KIC 5474065       & KIC 9726699\\
flare enery         & flare rate        & flare rate        & flare rate\\
 (ergs)             & (days)            & (days)            & (days) \\
\hline
$10^{30}$           & 0.09           &                   & 0.3\\ 
$10^{31}$           & 0.29           & 0.2               & 0.6\\
$10^{32}$           & 1.5            & 8.7                 & 117 \\
\hline
\end{tabular}
\end{center}
\label{flare-rate}
\end{table}

\section{Conclusions}

We present {\kep} short cadence observations of the M4 V star {\src}
which has a rotation period of 2.47 days.  It shows two high amplitude
short duration flares ($\Delta F/F>$0.4 which have integrated
energies of $\sim7\times10^{32}$ ergs. Additional flares energies as
low as $\sim1\times10^{32}$ ergs are also seen. We compare the flare
rate of a second M4 V star {\srctwo} which is more than 5 mag brighter
than {\src} and has a more rapid rotation period of 0.60
days. Compared to {\src}, {\srctwo} does not show such high amplitude
flares but since the {\kep} data has a higher signal to noise, it
allows us to detect many short duration, low energy flares reaching
energies as low as $\sim10^{30}$ ergs. Although the effect of flares
with energies of $10^{34}$ ergs on the atmosphere's of exoplanets in
the habitable zone have been investigated, it is of great interest to
determine what effect the presence of many numbers of lower energy
events will have on exo-planet atmospheres.

\section{Acknowledgements}

This paper includes data collected by the Kepler mission. Funding for
the Kepler mission is provided by the NASA Science Mission
Directorate. Some of the data presented in this paper were obtained
from the Mikulski Archive for Space Telescopes (MAST). STScI is
operated by the Association of Universities for Research in Astronomy,
Inc., under NASA contract NAS5-26555. Support for MAST for non-HST
data is provided by the NASA Office of Space Science via grant
NNX09AF08G and by other grants and contracts. Observations were also
made with the Gran Telescopio Canarias (GTC), installed in the Spanish
Observatorio del Roque de los Muchachos of the Instituto de
Astrofísica de Canarias, in the island of La Palma. Armagh Observatory
is supported by the Northern Ireland Government through the Dept
Culture, Arts and Leisure. We thank the referee for a constructive and
helpful report.

\end{document}